\documentclass{article}

\usepackage{arxiv}
\usepackage[super]{nth}
\usepackage{subcaption}
\usepackage{textcomp}
\usepackage[utf8]{inputenc} 
\usepackage[T1]{fontenc}    
\usepackage{hyperref}       
\usepackage{url}            
\usepackage{booktabs}       
\usepackage{amsfonts}       
\usepackage{nicefrac}       
\usepackage{amsmath}
\usepackage{microtype}      
\usepackage{graphicx}
\title{News and the city: understanding online press consumption patterns through mobile data}

\author{
  Salvatore Vilella \\
  ISI Foundation \\ \& \\
  Department of Computer Science \\ University of Turin\\
  Turin, Italy \\
  \texttt{salvatore.vilella@unito.it} \\
  \And
  Daniela Paolotti \\
  ISI Foundation \\ 
  Turin, Italy \\
  \texttt{daniela.paolotti@isi.it} \\
 \And
 Giancarlo Ruffo \\
  Department of Computer Science \\ University of Turin\\
  Turin, Italy \\
  \texttt{ruffo@di.unito.it} \\
  \And
  Leo Ferres \\
  Data Science Institute\\
  Universidad Del Desarrollo\\
  Santiago de Chile, Chile \\
  \texttt{lferres@udd.cl} \\
}

\begin{document}
\maketitle




\begin{abstract} 
	
The always increasing mobile connectivity affects every aspect of our daily lives, including how and when we keep ourselves informed and consult news media. By studying a DPI (\textit{deep packet inspection}) dataset, provided by one of the major Chilean telecommunication companies, we investigate how different cohorts of the population of Santiago De Chile consume news media content through their smartphones. We find that some socio-demographic attributes are highly associated to specific news media consumption patterns. In particular, education and age play a significant role in shaping the consumers behaviour even in the digital context, in agreement with a large body of literature on off-line media distribution channels.

This paper has been accepted for publication on the SpringerOpen journal EPJ Data Science, and is available at \url{https://doi.org/10.1140/epjds/s13688-020-00228-9}.
\end{abstract}



%




\section{Introduction}\label{intro}
{
Internet, the World Wide Web and, more recently, the pervasiveness of mobile technologies have radically transformed the way individuals consume cultural content. One of the areas that was impacted the most by new forms of digital media is journalism: how newspapers are accessed, how news are consumed. The abundance and diversity of topics have forever been altered \cite{alejandro2010journalism}. Gaining a better understanding of how different population groups are using and benefiting from on-line news services is now even more important. By leveraging mobile information, it is possible to map the variability of news consumption patterns of the population onto socio-demographic features such as the education level, age, income etc.

In this work, we want to explore the association between social inequality and news consumption in the context of digital news outlets. We focus on Chile, a country that in the past three decades has experienced an incredibly rapid economic growth along with a deep political transformation (from a dictatorship to a democracy). Exactly for these reasons, the case of Chile is particularly interesting. It is currently among the countries with larger social disparity and yet, according to the \textit{Newzoo's Global Mobile Market Report}, the penetration of digital technologies remains the highest in Latin America. The expected effect is that easier access to digital news outlet should “democratize” the news consumption. Indeed, it has been found that newspapers and magazines have lost their elite character, diversifying in order to adapt to popular preferences throughout the 20th century while books remain a marker of ‘elite culture’, and while status is not an important determinant of magazine and newspaper reading, education and income are \cite{torche2007social}. 

News consumption patterns have been increasingly studied in the last decade, especially now that digital media have started providing a huge variety of new platforms that facilitate - and, on the other hand, makes more complex - the fruition of news. Access to news outlets from mobile phones, for example, is now more fragmented and shallower with respect to traditional media \cite{molyneux2018mobile}. Technology also drives the consumption of content in terms of continuity of access ("anywhere, anytime") \cite{boczkowski2018news}. Previous literature \cite{lindell2018distinction} has also shown how individual choices with respect to news consumption can be influenced by the social status of the users. A strong and systematic association between status and newspaper readership has previously been established \cite{chan2007social}. Influence of education, income and age on newspaper consumption and platform preferences has also been explored \cite{anderson2018influence}. 

The role of education is particularly critical. Researchers have found education to be a strong predictor of media behavior \cite{self1988perceived}. It is also positively associated with general news exposure \cite{poindexter2001revisiting}.
In some context, news consumption is more unequally distributed than income is, with greater social inequality in online news consumption than in off line news consumption \cite{kalogeropoulos2017social}. Within the particular context of Chile, news media have already been studied to outline, for example, their ownership structures or their political bias and to understand the effects of certain press manipulations by owners of the media outlet \cite{bahamonde2018power}.

The contribution of this work is two-fold: first, we explore the association between socio-demographic attributes and news consumption in the Chilean context, in line with the literature we just described (that usually based the analyses in single-nation frameworks, like the US or Argentina) and our own previous work \cite{bahamonde2018power,elejalde2018nature,elejaldequant,elejaldetargetting}. Second, we rely upon a unconventional and unique data source: mobile phone data. Traditionally, this topic has been investigated using surveys among various population groups (\cite{rodrigues2018news,edgerly2018parents, fletcher2017paying,matthijs2019unpacking}, just to mention a few).  We believe that mobile technology can provide new insights that complement traditional methods of data collection. Indeed, many diverse research questions have been addressed using mobile communication data: to elaborate indices of the economic development of a region \cite{mao2015quantifying}, to study traffic flows, urban human mobility, social mixing phenomena \cite{gonzalez2008understanding, calabrese2011real, calabrese2015urban, iqbal2014development, graells2017effect, beiro2018shopping} as well as epidemic spreading on fine spatial and temporal scales \cite{wesolowski2016connecting}. We believe that such a data source could add interesting new insights also in the context of this research area, besides the fact that these methods are less expensive,  less time  consuming,  and  they scale efficiently.

In the following sections we present the results of a study on users accessing on-line news media in the city of Santiago de Chile (SCL) through mobile devices (using DPI (\textit{deep packet inspection}) data. The focus is on the time window that spans from the \nth{6} of July 2016 to the \nth{2} of August 2016. In order to characterize each area by the socio-demographic information of its residents, we used the 2017 census data (see Sec. \ref{census}). We find that some socio-demographic attributes are highly associated to specific news media consumption patterns. In particular, education and age play a significant role in shaping the consumers behaviour even in the digital context, in agreement with a large body of literature on off-line media distribution channels. }

\section{Methods}\label{methods}

\subsection{Datasets and data pre-processing}\label{datasets}
In this section, we present
a description of the data used and of the procedures that we followed for data 
cleaning and pre-processing.

\subsubsection{Census data} \label{census}

To characterize Chile socio-economically, we used the official 2017
Chilean census\footnote{Released by the \textit{Instituto Naciónal de
		Estadística (INE) See \url{https://www.censo2017.cl}}}, surveying
17,574,003 people (51.1\% females). Geo-politically, from largest to
smallest, Chile consists of ``regions'', the largest administrative
division, followed by ``comunas'' (similar to counties in the United
States), census districts, census zones, and finally blocks, the
finest level of geographical granularity. The public dataset is
available at the level of blocks. Here, we focus on the Metropolitan
Region of Santiago de Chile (RM, for short). The RM consists of 52
``comunas'' for a total of 7,112,808 individuals (51.3\% females),
accounting for about 40\% of the total population of Chile. In this
work we correlate census information with data from mobile web traffic
records. Since minors (persons whose age is  $< 18$)
cannot sign mobile phone subscriptions, we decided
to be conservative and only work with adults in the census to have a
better match between the sample populations of the mobile data and
census data sets, ending up with 5,450,592 individuals in the census
and 2,455,148 unique users in the general phone dataset.

The 2017 Census contains a great deal of information on several
aspects of the socio-economic composition of the Chilean
population. It has the issue that
it is only an abridged version of the usual census surveys, since a more comprehensive one will 
be released in 2022. Given the particular structure
of the questionnaire, we were able to perform a manual selection of the 
census features. Indeed, the vast majority of the questions in the survey
revolve around a few topics (e.g., on top of the basic question "\textit{Do you consider yourself
	as a member of a native minority?}" there are other additional questions to distinguish 
between the different native minorities). We thus decided to select our features only among
the following basic socio-demographic information:

\begin{itemize}
	\item Age;
	\item \textit{Escolaridad} (years of formal education attained);
	\item Student status (whether an individual is still studying or not);
	\item Membership to a native population.
\end{itemize}

\begin{figure}[h!]
\centering
	\includegraphics[scale=0.4]{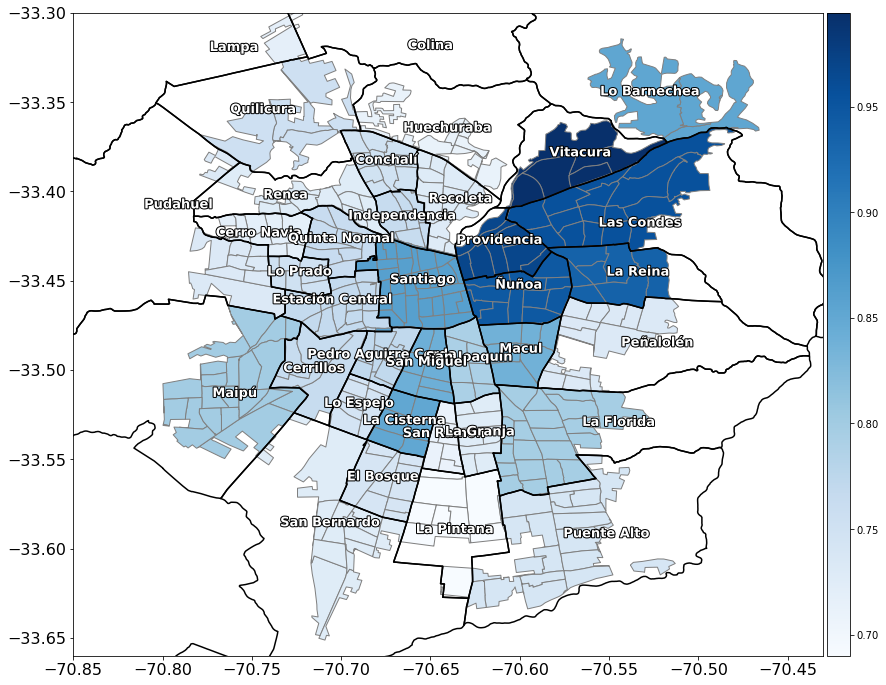}
	\caption{HDI distribution for the urban area of the \textit{Region Metropolitana de Santiago}
	}\label{HDI} 
\end{figure}

The census does not report on individuals' income directly. Thus,
information about the economic situation had to be inferred. To do so,
we chose \textit{escolaridad} as a proxy. In Chile, at least, there is
a well-known strong correlation between formal education and income
distribution and inequality
\cite{gregorio2002education,jerrim2015income}. Moreover, the variable 
\textit{escolaridad} weighs heavily in the calculation of the Human
Development Index (HDI), which is a widely used indicator of wellness 
and quality of life, that in turn can be used to get a first, qualitative
measure of the soundness of our results (a map of Santiago HDI distribution
at the level of municipalities can be seen in Fig. \ref{HDI}).

Census data can be aggregated at the different levels already
mentioned: region, comuna, districts, zones and blocks. As we go into
more granularity, census information becomes less specific in order to
avoid identification and preserve privacy of individuals. We have chosen
the level of {\em district} as a good trade-off between granularity,
privacy preservation and availability of information, see
Fig.\ref{CZimg}. Finally, each value was expressed as a percentage 
- or a mean, depending on the typology of the quantity - 
over the population of each district.

\begin{figure}[h!]
\centering
	\includegraphics[scale=0.4]{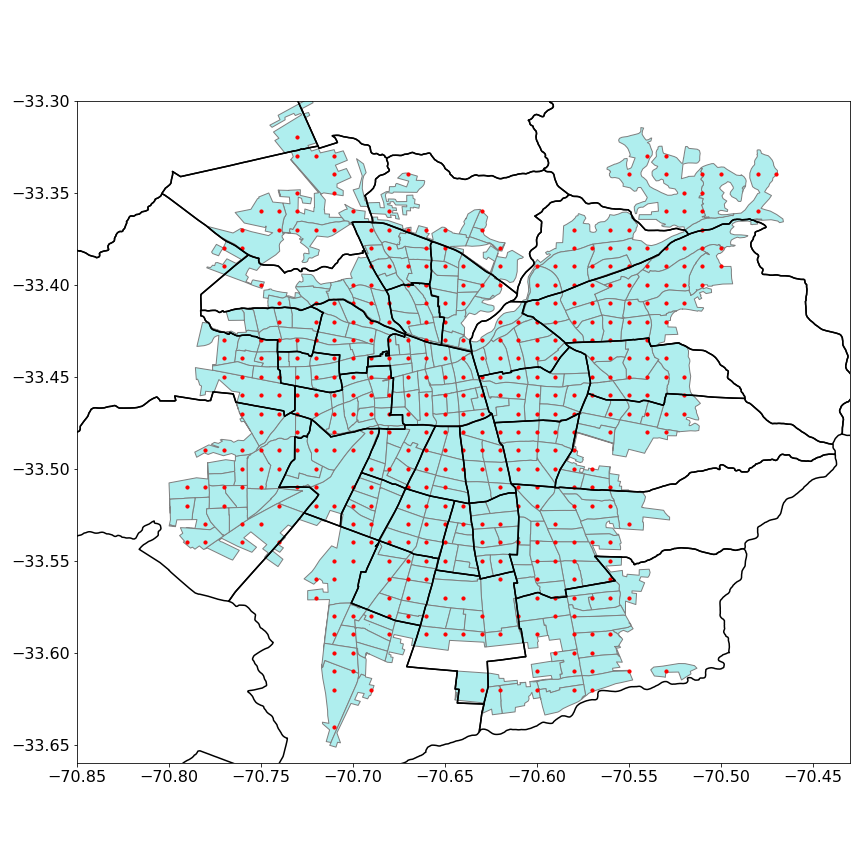}
	\caption{Census districts, comunas and \textit{dummy towers} in the urban area of Santiago.
		The census data is available at a level of \textit{census districts}, the small areas surrounded by white borders. In the Figure are also shown the boundaries of the \textit{comunas} (black lines), the administrative areas into which the city is divided. The red dots represent the dummy towers, obtained by clipping the coordinates of each antennas to the second decimal digit. All the traffic flow outgoing from each antenna was aggregated in the dummy towers to which it belongs.}label{CZimg} 
\end{figure}

Other than the provisos made above, the rest of the census dataset
was not modified or preprocessed in any other way.

\subsubsection{Building a mobile web connections data set} 
\label{dsusers}
We obtained a DPI (Deep Packet Inspection) data set from one of the largest company in terms of mobile subscriptions market share (around $30\%$) in Chile. {This data set is a record of internet connections from smartphones or any device that contains a SIM card and is subscribed to the telco network.} This dataset consists of almost a month (between the \nth{6} of
July 2016 and the \nth{2} of August 2016) of anonymized  \textit{events}. An event is defined as a connection of an individual device to an IP address through a cell tower (or antenna). In order to preserve privacy, information was aggregated by antenna and by hour, without any single user information, making it virtually impossible to de-anonymize news consumers.

Our {\tt DPI} data set includes the
\textit{number of unique users} that connected from an antenna to
an IP address at a certain hour (00, 01, 02... 23), as in the
following:

\begin{table}[h]
	\centering
	\scriptsize
	\begin{tabular}{lllllr}
		&antenna&date&hour&ip&usrs\\
		1&00000000 & 20160706 &11 & 200.12.26.117&2\\
		2&00000000&20160706&14&190.153.242.131&1\\
		3&00000000&20160706&14&200.12.20.11&1\\
		4&00000000&20160706&15&190.110.123.219&1\\
		... &... &... &... &... & ...
	\end{tabular}
	\caption{The first lines of the DPI dataset.}\label{snip_ds}
\end{table}

In the case of the small sample above, the first row of the raw
dataset tells us that from antenna {\tt 00000000}, on July 6, 2016, { there were two unique (i.e. distinct) users visiting 
{\tt 200.12.26.117} at 11 in the morning.}

The urban area of Santiago de Chile contains about 15,000 cell phone signal receivers, called {\em antennas}. Most outdoor antennas are placed on top of poles called cell phone {\em towers}. There are about 1,700 towers in our area of interest, with an average of 10 antennas per tower. There exist also standalone indoor antennas (sometimes called “small cells”), usually attached to indoor walls of large spaces such as hospitals, malls and public buildings. We have information on the exact latitude and longitude of the different {\em towers} and indoor antennas, allowing us to geo-reference our analyses. Abusing terminology, we call every unique geo-located cell phone receiver an ``antenna’’, this may be a tower with dozens of antennas on it, or just one single indoor antenna in a hospital. In high demand areas, such as the financial districts or industrial complexes, there area tens of towers with tens of antennas (plus indoor antennas) within a small area of a few tens of meters. Which antennas a user connects to depends on many factors, including distance, antenna demand (many devices may be connected to that antenna already), power, azimuth, among others. If the user moves a few meters, this could mean triggering the network to assign her a new antenna every few minutes. For that reason, to make the results more stable, in this paper, we group together all the antennas within a 1.1km radius, obtaining a grid of about 700 points. We call these ``dummy'' antennas (the red dots in Fig. \ref{CZimg}). ``Dummy'' towers are the result of clipping the full latitude and longitude  to two decimals, so that instead of having the ``real'' antennas at high-precision (and much finer level of granularity), we now have a ``dummy'' antenna at the center of the 1.1km2 square cell. These antennas should be taken as our ``sensors'', the finest level of granularity we have access to, which will also be mapped to the census, as we will see later.

Our dataset is a part of the complete \textit{deep packet inspection}
dataset of the telco provider. This subset contains only requests to those
IP addresses that belong to some kind of news media outlet. In order
to navigate among these outlets, we used a curated list of news 
organizations analyzed in a previous work \cite{bahamonde2018power}. 
These are around 400 news outlet accounts
\footnote{\url{https://bit.ly/2Ukpbyl}}, twenty six outlets for which
we knew their economic and political bias \cite{elejalde2018nature}. 

The process of identification of each IP address in order to associate
the name of a website was not straightforward. There was often a
many-to-one relationship between websites and IPs; for example, there
are clusters of two or more websites that share the same IP. This is a
critical issue, since in the dataset we can only see IP addresses,
without any DNS reverse lookup. In any case, most of the websites that
share the same IP belong to the same owner. This allows us to label
each IP by its owner: this way we lose knowledge about the individual
news outlet, that we are often unable to identify, but we still keep a
satisfying amount of information by creating a unique matching between
each IP and the editorial group (EG) that owns it. Thus, whenever
needed, the EG was considered instead of the individual news
media. Also, as shown in \cite{bahamonde2018power}, the power
structure in Chilean news media is strongly biased towards very few
groups that share an identifiable editorial line. The only newspaper
outside the above list is \textit{The Clinic}, a Chilean satirical
newspaper that is usually identified as \textit{leftist}, which we added in order to
cover the political spectrum as widely as possible, and use it as a
baseline for informational extremes (``El Mercurio'' would be
right-conservative, and ``The Clinic'' would be left-liberal). The
complete list of news outlets examined is shown in Table
\ref{npslist}.

{This selection is useful for various reasons: it allows us to remove some noise
caused by the outliers, i.e. not-so-popular news media that display a very low number of connections, and, 
most importantly, it helps us focusing on a narrow but significant range of news outlets, for which we 
already have a quantitative knowledge of their political bias and of their popularity. Incidentally, with this filtering
we were able to maintain in our analysis BioBioChile, El Mercurio and Cooperativa, that are the top 3 most visited outlets in our dataset, accounting for the vast majority of the volume of web traffic., thus not losing too much information.}

\begin{table}
	\centering
	\caption{List of news outlets (or editorial groups) examined in the {\tt DPI} data set. }\label{npslist}
	\begin{tabular}{c|c}
		\toprule
		\textbf{List of the news outlets examined}& {\textbf{Number of unique users } }\\ \midrule
		BioBioChile               &    1702768              \\
		El Mercurio editorial group       &    815152     \\
		Cooperativa     & 639120                           \\
		AdnRadioChile        & 465149                      \\
		The Clinic &    337048                       \\
		Tele 13       & 188409                             \\
		Diario Financiero       & 103450                    \\
		Publimetro Chile  & 54478                     \\
		\bottomrule
	\end{tabular}
\end{table}

{After filtering by news outlet and considering only the urban area of Santiago de Chile, we have $4313964$ events.}

\begin{figure}[h!]
\centering

	\includegraphics[scale=0.4]{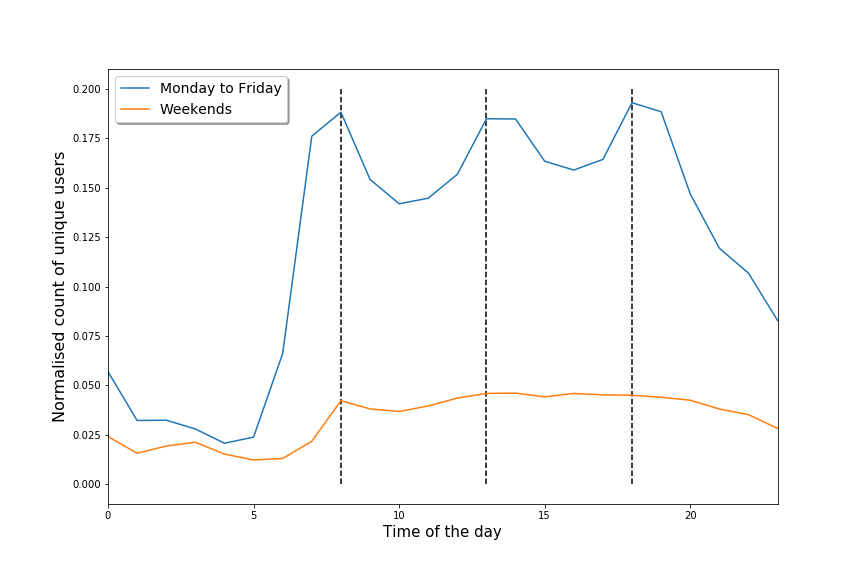}
	\caption{{Comparison between connections on weekends and working days. Values are normalized over the population of the area from where the connection is made. The dashed lines correspond to commuting hours: it is evident how during the weekends the effect of the \textit{floating population} are strongly attenuated.}}\label{trafficcompared} 
\end{figure}

Finally, since we are crossing mobile data with census, which contains 
socio-demographic information about the \textit{residents} of 
a certain area, we need to take into account the phenomenon 
of the \textit{floating population}, i.e. people moving from one 
place to another, especially at commuting hours and during
working days. This matter has been extensively 
studied, and there are study cases set up in Santiago
that show exactly how the city is affected by the phenomenon
(and how mobile data could be used to understand urban mobility,
see for example \cite{graells2016day}). 
To tackle this issue, we examined the temporal patterns of
the connections, illustrated in Fig.\ref{trafficcompared}. By comparing 
the trends between the weekends 
(Saturdays and Sundays) and the working days (Mondays to Fridays)
we can easily notice the circadian rhythms of the city: the peak in connections
starts when people wake up, continues when they commute
to work, rises again at lunch time and finally when they go back home at 6 pm.
This last peak is what differentiates the working days from the weekends:
on Saturdays and Sundays, no afternoon peak can be observed, with a smooth
decline of the connections towards night hours instead. The typical effects of the 
floating population phenomenon seem much attenuated during the weekends, {and the volume of connections seems to
remain almost constant throughout the whole day.}
{The correlation between the number of unique users and the number 
of residents in each comuna is rather high, with a Pearson coefficient P=0.75 $(\text{p-value}=
1.85e-07)$. This means that the geographical distribution of users accessing news media websites from mobile is a good proxy for the geographical distribution of the population.}
{There is a clear trade-off in our choice of 
considering only the connections at the weekends: on the one hand we 
reduce our data to almost one quarter of the initial volume, i.e. $\sim 1$ M events, but on the
other hand we are able to address the issue of the 
\textit{floating population}, at least partially 
reducing the noise caused by this phenomenon.}

\subsection{Ethical considerations}
Working with mobile data
could raise privacy concerns. Of course, DPI data was handed out by the mobile provider 
already anonymized and grouped per tower: data appeared as shown in Table \ref{snip_ds}, displaying
only the number of unique devices connected to a certain antenna, without any knowledge about the
identity or the profile of the customer. Therefore, we work with highly aggregated data
and, for good measure, the resolution was then made even coarser, as described in Sec. \ref{dsusers}.
The same applies to census data:
even though they are {already} anonymized and publicly available, the research was not carried out
at the finest possible granularity, in order to have a good balance between available amount of information and 
privacy preservation. {Our concern in relation to privacy stems from the possibility of cross-referencing raw DPI data with census information which could have unforeseen consequences and result in potential de-anonymization.
}

No attempt was made to infer statistics about individuals: the nature of DPI itself prevents this from
happening. All the results are \textit{general} trends of \textit{very large} groups of residents - between 
hundreds of thousands and millions of people.

\subsection{Mapping census data to districts}

\begin{figure}[h!]
\centering

	\includegraphics[scale=0.4]{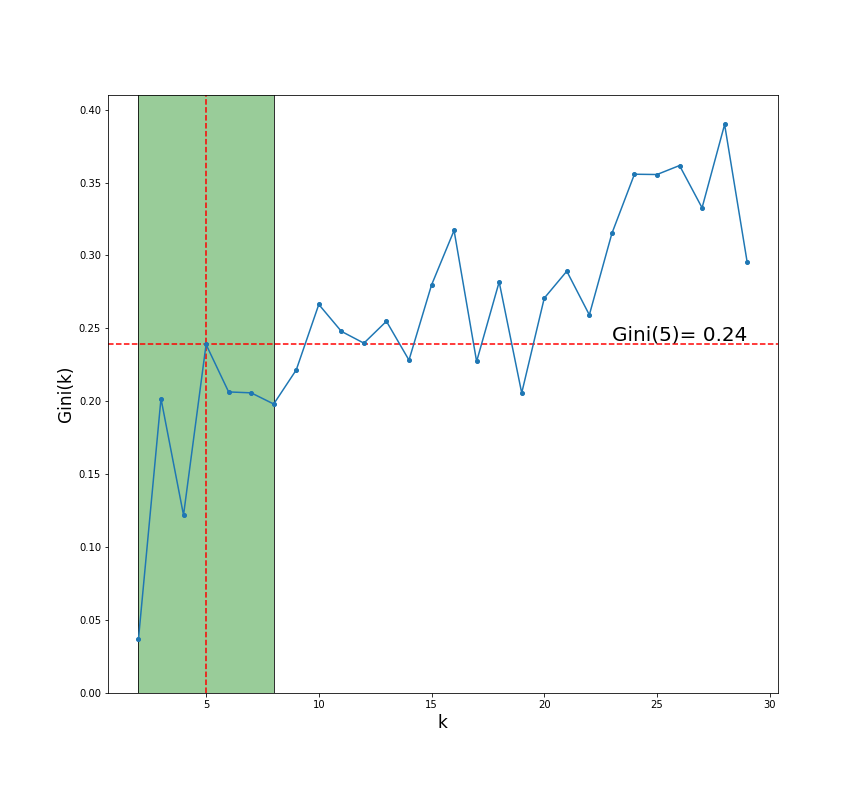}
	\caption{Gini coefficient for different values of $K$. $K=5$ was chosen as it maximizes the coefficient. The green area refers to the values of $K$ that we
		kept into consideration for the final choice. }\label{Gini} 
\end{figure}

All the connections to the websites are grouped by hour and by 
antenna: this provides with a good approximation of the users' position in
the city. Our goal is to study the connections based on the census
features of the areas from where they are originated. Thus, we want to 
assign to each census district (CD) a label that helps us distinguishing 
between different segments of census. To estimate the most appropriate number of groups (and, therefore, labels),
we cluster the CDs 
based on the census features that we selected earlier. We use a 
k-means algorithm, running the procedure several times for $2 \leq k \leq 30$.
Since we need a clear and easily interpretable grouping, we limit our
choice to $3 \leq k \leq 8$: having only 2 census groups would not be
enough for such an analysis, while more than 8 would be hard to 
interpret. {The values are standardized, and the algorithm was initialized with a random seed}. 
For each $k$ we compute the Gini coefficient of the distribution 
of the population (Fig. \ref{Gini}): a low Gini represents a situation of order,
in which the population is almost equally distributed across
all the clusters, while a high value is retrieved for an heterogeneous distribution throughout the different groups.

The final value, $k=5$, is chosen as it is the value that maximizes the Gini 
coefficient and, thus, the heterogeneity of the distribution of the population. This is based on the assumption that a very ordered situation, in
which we have almost the same number of people in each cluster, would be non realistic.

By setting $k=5$ and inspecting
the average values of the features for each group (see Table \ref{clustab}), we observe an almost 
hierarchical relationship among the different clusters. This is particularly true
for the \textit{escolaridad} variable, which correlates very well with
the income, as mentioned above. The clusters were simply named \textit{K1-K2-K3-K4-K5}, with K1 being the wealthiest area, 
and all the others following in an ordinal fashion. {In light of this hierarchical relationship, we will denote these groups - or clusters - also as census \textit{levels}.}

\begin{table}
	\centering
	\begin{tabular}{ccccc}
		\toprule
		\textbf{Cluster} &   Mean age & Avg years of schooling & \% of students & \% of people of indigenous ethnicity \\
		\midrule
		\textbf{K1} &       $46.25\pm0.27$ &       $16.91\pm0.11$ &    $0.154\pm0.003$ & $0.047\pm0.002$ \\
		\textbf{K2} &       $38.78\pm0.58$ &       $16.50\pm0.17$ &        $0.179\pm0.006$ &    $0.075\pm0.004$ \\
		\textbf{K3} &       $42.05\pm0.21$ &       $14.65\pm0.12$ &        $0.138\pm0.002$ &  $0.096\pm0.002$ \\
		\textbf{K4} &      $46.36\pm0.13$ &       $14.30\pm0.09$ &        $0.122\pm0.001$ &   $0.097\pm0.001$ \\
		\textbf{K5} & $44.62\pm0.18$ &  $12.86\pm0.08$ &   $0.106\pm0.001$ &   $0.131\pm0.002$ \\
		\bottomrule
	\end{tabular}

	\caption{Insights of the clusters. In these table are shown the mean values of the features for each cluster.}\label{clustab}
\end{table}

\begin{figure}[h!]
\centering

	\includegraphics[scale=0.4]{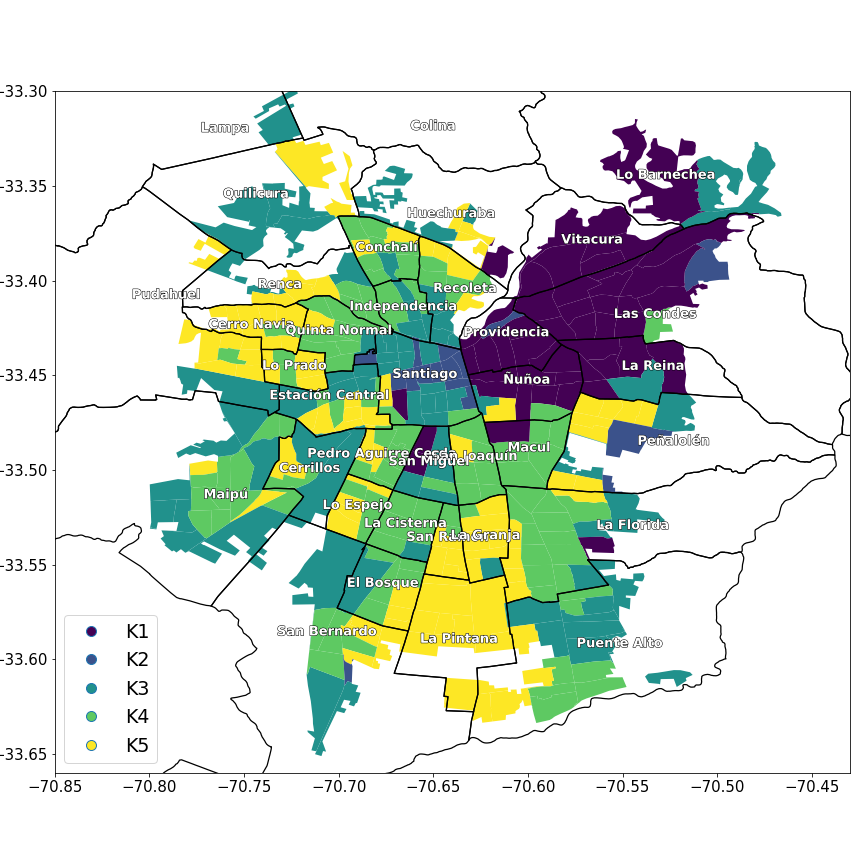}
	\caption{Result of the k-means clustering on the census districts.}\label{census_k} 
\end{figure}

The resulting map, shown in Fig. \ref{census_k}, resembles the distribution of the HDI shown in Fig.\ref{HDI}. In particular, 
the clustering procedure captures very well the presence of a very rich and highly segregated area (cluster K1) 
in the north-eastern part of the urban area, while the other census groups are distributed more heterogeneously 
throughout the city.

\subsubsection{Study of the global spatial autocorrelation}\label{moran}

\begin{figure}[h!]
\centering

	\includegraphics[scale=0.6]{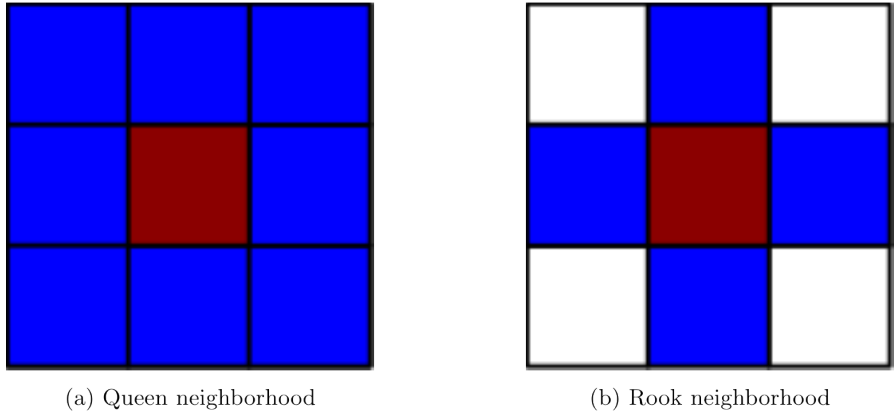}
	\caption{Types of neighborhood for the spatial weight matrix.}\label{neigh_a} 
\end{figure}

As further evidence of the segregation that emerges from the k-means algorithm, 
we dig deeper into the spatial distribution of the census features. To do this, we measure Moran's
Index \cite{moran1950notes} on census data and test the global
spatial autocorrelation of the features against a random null model. Moran's I for a
variable $y$ measured over $n$ spatial units is defined as
$$I=\frac{n}{S_{0}}\frac{\sum_{i}\sum_{j}z_{i}w_{ij}z_{j}}{\sum_{i}z_{i}^{2}}$$
where $w_{i,j}$ are the elements of a spatial weight matrix $W$,
$S_{0}=\sum_{i}\sum_{j}w_{ij}$ and $z_{i}=y_{i}-\bar{y}$ is the
deviation of \textit{y} from its mean value in the spatial unit
\textit{i}. The matrix $W$ is essential in spatial autocorrelation
analysis, since it provides the model with a measure of spatial
contiguity. The definition of spatial contiguity is usually specified
as a \textit{neighborhood} relationship between spatial units. Since
we are dealing with census districts, namely areas of the city, we decided
to use the Queen neighborhood (Fig. \ref{neigh_a}). { We believe that using
a Rook neighborhood in a context of urban spatial analysis
would be too limiting because, for each computation, we would not be considering the influence of the
areas in the corners of the spatial unit under examination}.
Thus, we defined as \textit{contiguous}
all those areas that surround the zone we are observing,  {considering sufficient that they only meet at corner vertexes \cite{o2014geographic}}: all those
areas interact, communicate and most likely influence one another.

We measured the global spatial autocorrelation for the census since we need 
to corroborate any evidence of
clustering resulting from the application of the k-means algorithm.
The Moran's I is an univariate measure of correlation, hence we
measured it for every census feature. All the results were then
compared with the Moran's I calculated for the case of a completely
random spatial distribution of the features.

In order to have a better understanding of the spatial autocorrelation and have a
\textit{more local} view, we examined the Moran scatterplot of the
data points for each feature. The analysis of the Moran scatterplot
was first introduced by Anselin \cite{anselin1993moran}, and it is
based on the interpretation of the Moran's Index as a coefficient of
an ordinary least squares (OLS) regression of the \textit{spatially lagged} variable on the
variable itself:

$$\sum_{j}w_{ij}z_{j}= a + Iz_{i}$$

where $\sum_{j}w_{ij}z_{j}$ is the \textit{spatial lag} of the
variable \textit{z} (which in turn is $y$ expressed as deviation from
the mean). Hence the spatial lag is a weighted average of the variable
$z$ over the neighbours of $i$. Note that the interpretation of $I$ as
a coefficient of an OLS regression is valid for any statistic that can
be expressed as a ratio between a quadratic form and the sum of the
squares \cite{anselin1993moran}, which is exactly how the Moran's I is defined.

Plotting the spatial lag of $z$ as a function of $z$ allows us to have
a local view of the spatial autocorrelation. The plot is divided into
four quadrants, going from the 1st quadrant of high values of both $z$
and the spatial lag (\textit{high-high correlation}) to the 4th
quadrant of high-low correlation, passing through the so-called
low-high and low-low correlation. A point being far up in the first
quadrant of the Moran scatterplot of the \textit{average schooling
	year} feature, represents an area that has a very high mean value of
schooling, and is surrounded by neighbours with similarly high values.

\begin{figure}[h!]
\centering

	\includegraphics[scale=0.35]{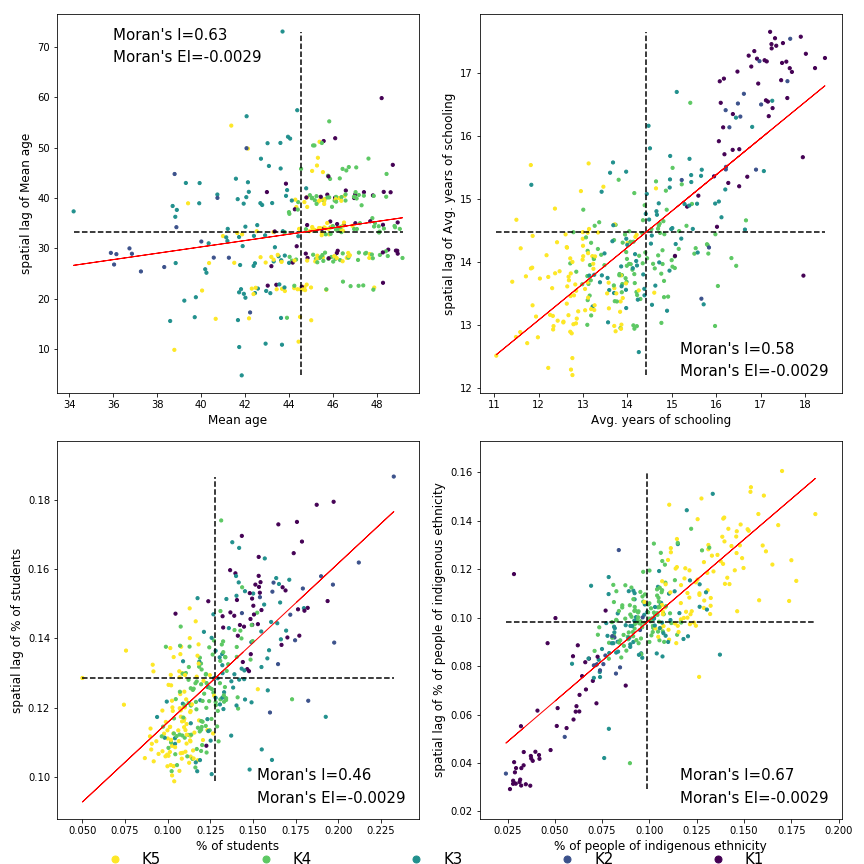}
	\caption{Moran scatterplot for the census features. It is clear how, for all the features but the mean age, the census segments (K1 and K5 in particular) are strongly segregated. We observe different patterns for the mean age, where we can see how K2 is shifted to the left, towards a lower average value of the age. }\label{moranscatcens} 
\end{figure}

As we can see in Fig. \ref{moranscatcens}, there is strong evidence of 
segregation for almost all the census features, in particular for the 
\textit{extreme} census segments K1 and K5 which, depending on the feature, 
are usually located way up near the bisector of the first and third quadrants.
The census segments in between, instead, display a far more mixed 
situation.

\section{Results and discussion}
\label{results}

\subsection{Geo-referenced analysis of the {\tt DPI} dataset}

The labeling of the census zones based on socio-demographic features finally enables us to proceed with a
geo-referenced analysis of the connections towards the selected news
outlets websites. The goal is to find differences in the consumption 
of news media content by areas of the city that have different socio-demographical
attributes. 

A list of the news outlets can be found in Table
\ref{npslist}: wherever we encountered the issues described in
Sec. \ref{dsusers}, we grouped the news outlets by owner and considered
the resulting group as an individual entity. This is the case of what
happened with all the outlets belonging to the \textit{El Mercurio}
editorial group, which we included in our final list as a single
entry.

\begin{figure}[h!]
\centering

	\includegraphics[scale=0.4]{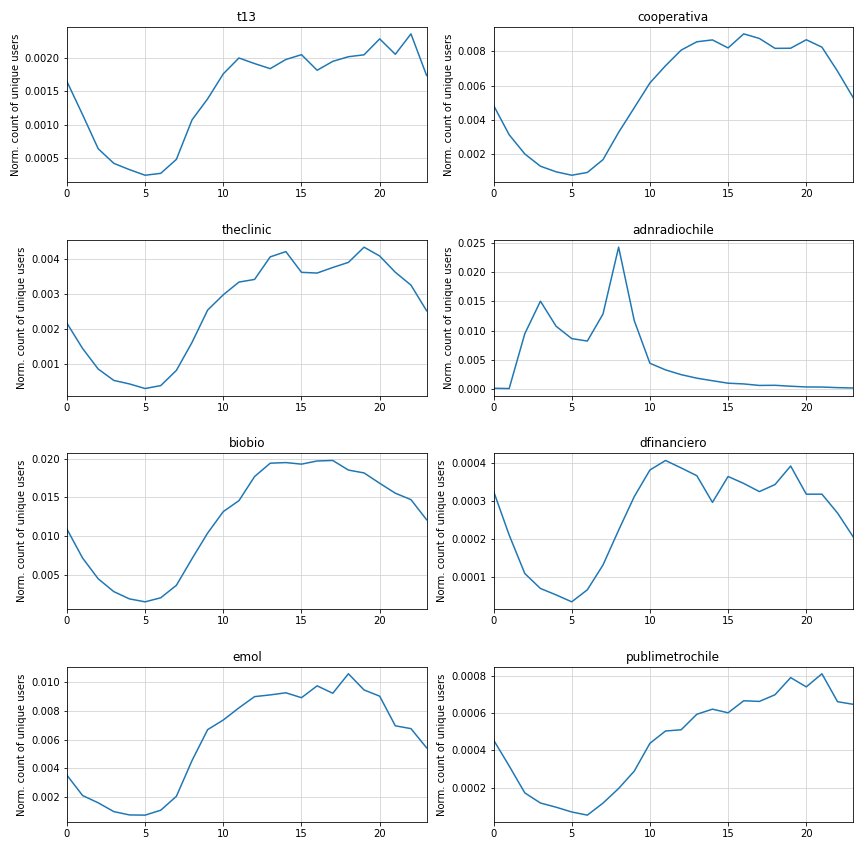}
	\caption{Unique users
		over the whole time window studied, detailed by hour and by news outlet. On the horizontal axis we have the time of the day, and on the vertical axis the normalized accesses. From these plots clearly emerges the circadian rhythm followed by the temporal patterns of most of the news outles. }\label{traffic_ind} 
\end{figure}

\begin{figure}[h!]
\centering

	\includegraphics[scale=0.4]{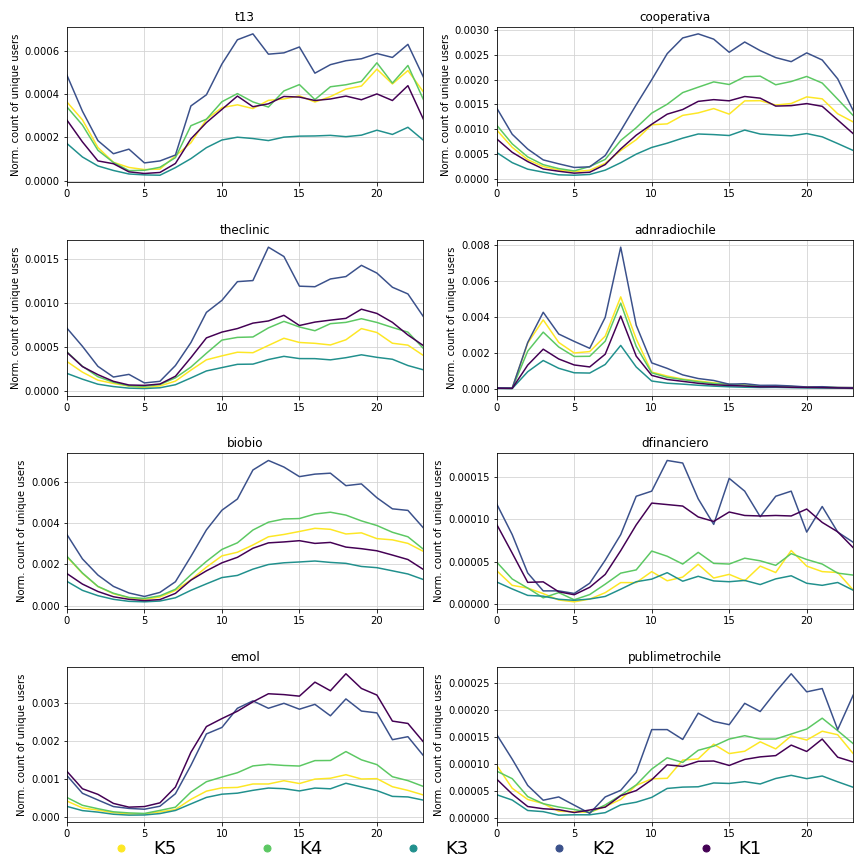}
	\caption{Unique users
		over the whole time window studied, detailed by hour, by news outlet and by cluster. On the horizontal axis we have the time of the day, and on the vertical axis the normalized number of users. These plots show us how the different clusters contribute to the web traffic of the individual news outlets. }\label{traffic} 
\end{figure}

\begin{figure}[h!]
\centering

	\includegraphics[scale=0.4]{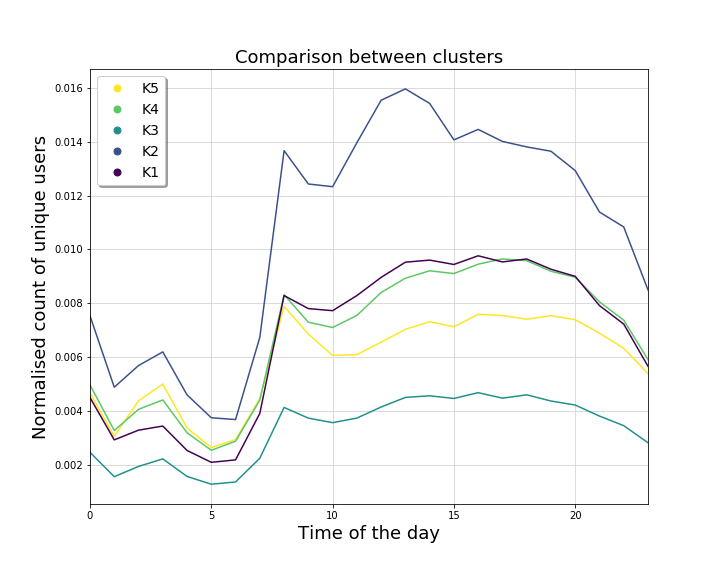}
	\caption{Total connections,
		detailed per hour and disaggregated by census segment of the area of origin.}\label{trafficaggregated} 
\end{figure}

In Figure \ref{traffic} we plot, for each News Outlet (NO), 
the number of unique users in each 
CD, normalized over the number of residents of the cluster the CD belongs
to. Fig. \ref{trafficaggregated} shows the same quantity considering all the 
NOs together, i.e. the total {amount of users connecting to} news outlets websites during the 
examined weekends of July and August 2016. 

Before analyzing these plots, it is worth going back to Table \ref{clustab}.
The grouping into 5 clusters portrays a very clear division between the first
two census segments and the remaining three. Indeed, groups K1 and K2 
display a very high average value of schooling years ($>16$), with group K2 
being composed of much younger individuals and a higher percentage 
of students. Groups K3, K4 and K5 instead are more comparable with each other,
displaying a progressively lower value of schooling years and percentage of students, 
and a higher value of people of indigenous ethnicity. As pointed out in 
Sec. \ref{moran}, these three groups are spatially mixed, while K1 and K2 are
segregated in the north-eastern and, to a lower extent, in the central part of
the urban area of the \textit{Region Metropolitana de Santiago}.

{In light of} this grouping, we can inspect the results shown in the plots. From
Fig. \ref{trafficaggregated} we can gather the first, clear evidence: group K2
completely dominate the chart. This means that the majority of the connections 
are originated in those areas that are characterized by a high percentage of young 
and highly educated residents. As said, for each cluster the connections have been 
normalized over the number of people in it, meaning that we are looking at 
a measure of the \textit{activity} of a cluster with respect to its size. This dominance
of the group K2 and, in smaller measure, of K1, is confirmed also by 
computing the Pearson correlation between the number of connections and the number
of residents in each cluster. We get P=-0.92 $(\text{p-value}=0.029)$, a very high 
anti-correlation, meaning that - in proportion to their size - the smallest clusters 
show the highest activity. Indeed, the smallest clusters are K2 (with only 255699 people)
and K1 (682053 inhabitants), while the remaining three all have more than 1M residents each. 

We can also analyze the same trends for
the individual news media websites. From Fig. \ref{traffic} we can see that the areas 
belonging to group K2 are again on top in the majority of the plots. The only exception
is for the websites belonging to \textit{El Mercurio}, a right wing
editorial group, which is most accessed by those areas belonging
to the K1 segment. {Similar} numbers between K1 and K2 can be found 
also for the website of \textit{Diario Financiero}, a financial newspaper.
As for the other census groups no clear pattern can be identified, but still
it seems pretty clear that, both in Fig. \ref{traffic} and \ref{trafficaggregated}, 
group K3 is always the least active. In general, segments K3, K4 and K5 display an
activity comparable to that of groups K1 and K2 only for those news media, 
like \textit{Biobio}, \textit{Cooperativa} or \textit{Tele13}, that can be  classified as 
containers of \textit{generic} news and information, unlike \textit{Diario Financiero}, 
which is highly specialised on financial matters, or  \textit{El Mercurio}, which is openly
politically oriented.

As expected (\cite{de2016death,graells2016sensing,graells2017effect}), the circadian rhythm of 
the population can still be seen from these figures, even though 
to a lower extent than the same analysis performed on the weekdays (Fig. \ref{trafficcompared}), 
as explained in Sec. \ref{dsusers}. The decline in connections
towards night hours is smooth, and the trends are very similar across all the 
census segments.

\subsubsection{Addressing the limitations}
This method presents some limitations. First of all, we acknowledge a possible
bias in our dataset: while census is representative of the whole
population, mobile data accounts only for the customers of the
service provider, which can be biased towards a certain segment of the
population depending on multiple factors (typologies of the proposed
contracts, marketing strategies, etc.). Nonetheless, bearing in mind
that this provider alone owns more than the 37\% of the marketshare, a
correlation check between our dataset and the number of residents in
each census zone gave us satisfactory results, as pointed out in Sec. \ref{dsusers}. 
Another influence on
the results could come from the diverse penetration rate of WiFi
technologies among the various census groups. One of our basic
assumptions is that people, during weekends, access the web mostly
from their houses or from nearby locations. This means that, especially
in their houses but also in some public places, people could be surfing the
web via WiFi instead of using their own mobile data. This, again, strongly
depends on the specific country and on its telco market, and it could 
also be related to the fact that -
in general - WiFi is more popular among those individuals that have,
for example, a bank account (sometimes needed even just to get a
broadband contract in the first place) or at least a certain yearly
income. Thus, we acknowledge that these census groups, in particular
K1 and K2 segments according to
our clustering algorithm, could be slightly under-represented in our
dataset. Nonetheless, the positive correlation between the number of
unique accesses and the residents in each municipality is a
reassuring evidence with respect to this matter. Moreover, for the
particular case of Chile, the effects of this issue should be limited:
according to official data (Subtel), nearly 80\% of all internet
access is made from mobile phones \cite{report}.

{A further limitation of this study is data availability. We are under strict non-publication policies from the telco, and the data cannot leave the telco’s premises. This is to be expected given the novelty of the data sets we are working with, and their yet-to-be-understood privacy implications (this notwithstanding, please see the “Data Availability” section below)}

\section{Conclusions}

The main purpose of this work {is} to understand {whether individuals living in} different areas of the
city of Santiago de Chile {display different behaviors in terms of} access to {digital} news media content  {with respect to socio-demographic attributes}. In order to do so, we studied a record of a month (July-August 2016)
of anonymized and geo-referenced accesses to several websites of Chilean news media
through the cellphone network of a major telco provider. To reduce the effects of the \textit{floating population} phenomenon, we analyzed only a portion of the connections, narrowing the window to the weekends only.

By applying a k-means algorithm on official Chilean census data, in order to group the population
into 5 different census levels, we find that the wealthiest areas of the city, K1 and K2, are very similar in terms
of \textit{years of schooling}, with K2 members being - on average - much younger than K1. 
The other three clusters (K3-K4-K5) are indeed pretty far from the first two in terms 
of average education level, and far more spatially mixed: K1 and K2 are mostly grouped in the north-eastern quadrants of the city.

The news access patterns tell us that segment K2 is overall the most 
active group. {This means that the most highly educated and youngest are also the most informed, with respect to older people with similar education level. 
If we break down the number of access by individual media outlet, we obtain a more composite picture (Fig. \ref{trafficcompared}). K2 remains the most comprehensively informed with respect to the diversity of media outlets; on the other hand the ranking of the other groups can vary depending on the flavour of the news media.}

These findings confirm empirically that socio-demographic features have an influence on the consumption of news. Our results suggest that the
consumption of news media content does not always necessarily increase with the education level of the user. While it is true that the highly educated are those who access news media websites the most, the opposite cannot be inferred: indeed group K3, which includes all those areas of the city whose residents are averagely educated (see Table \ref{clustab}), is the cluster that displays the lowest activity.

The predominance of group K2 also tells us that age plays a key role in this context. Young generations are likely more comfortable in using mobile devices in their daily lives, and this has been also explored in the specific field of news consumption \cite{chan2015examining}.

In summary, although we found clear correlations between the consumption of online news media content and the socio-demographics of users, this relationship seems to be non-trivial. While the highly educated are the most eager users of mobile news media outlets, there are important clues that a group of well educated people, that we could identify as middle-class, show a lack of interest in accessing news media via mobile. This could have a significant impact on the civic engagement of a relevant share of the population, as well as on the economical and political life of a Country, not mentioning the media market itself.

There are positive correlations between the other two features - \textit{student status} and \textit{percentage of people of native ethnicity}. Therefore, a more detailed analysis of the ties between these features and the consumption patterns of the individual news outlets - namely, the habits and preferences of students and native minorities - could be of much interest in the context of studying inequalities in the accesses between different cohorts of the population, and it is left as future work.

{We believe that this work could be useful to shed some light on how digital platforms can contribute to the already complex interplay between socio-demographic characteristics of the population and the news consumption behaviour. Additional studies with more comprehensive datasets in other social contexts are needed, to assess the full impact of digital technologies on this topic, as well as to understand if these findings are highly Country-specific or if there is a shared worldwide-trend.}


\section*{Competing interests}
  The authors declare that they have no competing interests.

\section*{Author's contributions}
    Conceptualization: LF, DP, GR, SV; Data curation:  SV, LF; Formal analysis and Methodology: DP, GR, LF, SV; Writing: SV, LF, DP, GR.

\section*{Acknowledgements}
Daniela Paolotti and Salvatore Vilella acknowledge support from the Lagrange Project of the Institute for Scientific Interchange Foundation (ISI Foundation) funded by Fondazione Cassa di Risparmio di Torino (Fondazione CRT). LF acknowledges financial support
from Movistar - Telefónica Chile, the Chilean government initiative
CORFO 13CEE2-21592 (2013-21592-1-INNOVA\_ PRODUCCION2013-21592-1),
Conicyt PAI Networks (REDES170151) ``Geo - Temporal factors in disease
spreading and prevention in Chile'', and Project PLU180009, Tenth
``Fondo de Estudios sobre el Pluralismo en el Sistema Informativo
Nacional'', 2018, `` Geo-Temporal Access to Chilean news outlets using
digitales traces''.
\section*{Data availability}
2017 Chilean Census data is available at \url{https://www.censo2017.cl}. DPI dataset was provided by the mobile telecommunication company, and 
therefore is not publicly available. However, if any researcher would like to access/work on the data, the telco’s offices are open to collaboration through the authors of this paper. Please send an email to \verb+lferres@udd.cl+.


\bibliographystyle{unsrt}  
\bibliography{bmc_article.bib}  

\begin{thebibliography}{10}

\bibitem{alejandro2010journalism}
Jennifer Alejandro.
\newblock Journalism in the age of social media.
\newblock {\em Reuters Institute Fellowship Paper}, page~5, 2010.

\bibitem{torche2007social}
Florencia Torche.
\newblock Social status and cultural consumption: The case of reading in chile.
\newblock {\em poetics}, 35(2-3):70--92, 2007.

\bibitem{molyneux2018mobile}
Logan Molyneux.
\newblock Mobile news consumption: A habit of snacking.
\newblock {\em Digital Journalism}, 6(5):634--650, 2018.

\bibitem{boczkowski2018news}
Pablo~J Boczkowski, Eugenia Mitchelstein, and Mora Matassi.
\newblock “news comes across when i’m in a moment of leisure”:
  Understanding the practices of incidental news consumption on social media.
\newblock {\em New Media \& Society}, 20(10):3523--3539, 2018.

\bibitem{lindell2018distinction}
Johan Lindell.
\newblock Distinction recapped: Digital news repertoires in the class
  structure.
\newblock {\em New Media \& Society}, 20(8):3029--3049, 2018.

\bibitem{chan2007social}
Tak~Wing Chan and John~H Goldthorpe.
\newblock Social status and newspaper readership.
\newblock {\em American journal of sociology}, 112(4):1095--1134, 2007.

\bibitem{anderson2018influence}
Bryan Anderson.
\newblock Influence of education, income and age on newspaper use and platform
  preference.
\newblock {\em Elon Journal of Undergraduate Research in Communications},
  9(1):108--114, 2018.

\bibitem{self1988perceived}
Charles~C Self.
\newblock Perceived task of news report as a predictor of media choice.
\newblock {\em Journalism Quarterly}, 65(1):119--125, 1988.

\bibitem{poindexter2001revisiting}
Paula~M Poindexter and Maxwell~E McCombs.
\newblock Revisiting the civic duty to keep informed in the new media
  environment.
\newblock {\em Journalism \& Mass Communication Quarterly}, 78(1):113--126,
  2001.

\bibitem{kalogeropoulos2017social}
AK~Kalogeropoulos and Rasmus~Kleis Nielsen.
\newblock Social inequalities in news consumption.
\newblock 2017.

\bibitem{bahamonde2018power}
Jorge Bahamonde, Johan Bollen, Erick Elejalde, Leo Ferres, and Barbara Poblete.
\newblock Power structure in chilean news media.
\newblock {\em PloS one}, 13(6):e0197150, 2018.

\bibitem{elejalde2018nature}
Erick Elejalde, Leo Ferres, and Eelco Herder.
\newblock On the nature of real and perceived bias in the mainstream media.
\newblock {\em PloS one}, 13(3):e0193765, 2018.

\bibitem{elejaldequant}
Erick Elejalde, Leo Ferres, Eelco Herder, and Johan Bollen.
\newblock Quantifying the ecological diversity and health of online news.
\newblock {\em Journal of Computational Science}, 27:218 -- 226, 2018.

\bibitem{elejaldetargetting}
{Elejalde, Erick}, {Ferres, Leo}, and {Schifanella, Rossano}.
\newblock Understanding news outlets\'{} audience-targeting patterns.
\newblock {\em EPJ Data Sci.}, 8(1):16, 2019.

\bibitem{rodrigues2018news}
Usha~M Rodrigues and Yin Paradies.
\newblock News consumption habits of culturally diverse australians in the
  digital era: Implications for intercultural relations.
\newblock {\em Journal of Intercultural Communication Research}, 47(1):38--51,
  2018.

\bibitem{edgerly2018parents}
Stephanie Edgerly, Kjerstin Thorson, Esther Thorson, Emily~K Vraga, and Leticia
  Bode.
\newblock Do parents still model news consumption? socializing news use among
  adolescents in a multi-device world.
\newblock {\em new media \& society}, 20(4):1263--1281, 2018.

\bibitem{fletcher2017paying}
Richard Fletcher and Rasmus~Kleis Nielsen.
\newblock Paying for online news: A comparative analysis of six countries.
\newblock {\em Digital Journalism}, 5(9):1173--1191, 2017.

\bibitem{matthijs2019unpacking}
Koen Matthijs, David De~Coninck, Marlies Debrael, Leen d'Haenens, and Rozane
  De~Cock.
\newblock Unpacking attitudes on immigrants and refugees: a focus on household
  composition and news media consumption.
\newblock {\em Media and Communication}, 7(1):43--55, 2019.

\bibitem{mao2015quantifying}
Huina Mao, Xin Shuai, Yong-Yeol Ahn, and Johan Bollen.
\newblock Quantifying socio-economic indicators in developing countries from
  mobile phone communication data: applications to c{\^o}te d’ivoire.
\newblock {\em EPJ Data Science}, 4(1):15, 2015.

\bibitem{gonzalez2008understanding}
Marta~C Gonzalez, Cesar~A Hidalgo, and Albert-Laszlo Barabasi.
\newblock Understanding individual human mobility patterns.
\newblock {\em nature}, 453(7196):779, 2008.

\bibitem{calabrese2011real}
Francesco Calabrese, Massimo Colonna, Piero Lovisolo, Dario Parata, and Carlo
  Ratti.
\newblock Real-time urban monitoring using cell phones: A case study in rome.
\newblock {\em IEEE Transactions on Intelligent Transportation Systems},
  12(1):141--151, 2011.

\bibitem{calabrese2015urban}
Francesco Calabrese, Laura Ferrari, and Vincent~D Blondel.
\newblock Urban sensing using mobile phone network data: a survey of research.
\newblock {\em Acm computing surveys (csur)}, 47(2):25, 2015.

\bibitem{iqbal2014development}
Md~Shahadat Iqbal, Charisma~F Choudhury, Pu~Wang, and Marta~C Gonz{\'a}lez.
\newblock Development of origin--destination matrices using mobile phone call
  data.
\newblock {\em Transportation Research Part C: Emerging Technologies},
  40:63--74, 2014.

\bibitem{graells2017effect}
Eduardo Graells-Garrido, Leo Ferres, Diego Caro, and Loreto Bravo.
\newblock The effect of pok{\'e}mon go on the pulse of the city: a natural
  experiment.
\newblock {\em EPJ Data Science}, 6(1):23, 2017.

\bibitem{beiro2018shopping}
Mariano~G Beir{\'o}, Loreto Bravo, Diego Caro, Ciro Cattuto, Leo Ferres, and
  Eduardo Graells-Garrido.
\newblock Shopping mall attraction and social mixing at a city scale.
\newblock {\em EPJ Data Science}, 7(1):28, 2018.

\bibitem{wesolowski2016connecting}
Amy Wesolowski, Caroline~O Buckee, Kenth Eng{\o}-Monsen, and Charlotte
  Jessica~Eland Metcalf.
\newblock Connecting mobility to infectious diseases: the promise and limits of
  mobile phone data.
\newblock {\em The Journal of infectious diseases}, 214(suppl\_4):S414--S420,
  2016.

\bibitem{gregorio2002education}
Jose~De Gregorio and Jong-Wha Lee.
\newblock Education and income inequality: new evidence from cross-country
  data.
\newblock {\em Review of income and wealth}, 48(3):395--416, 2002.

\bibitem{jerrim2015income}
John Jerrim and Lindsey Macmillan.
\newblock Income inequality, intergenerational mobility, and the great gatsby
  curve: Is education the key?
\newblock {\em Social Forces}, 94(2):505--533, 2015.

\bibitem{graells2016day}
Eduardo Graells-Garrido and Diego Saez-Trumper.
\newblock A day of your days: estimating individual daily journeys using mobile
  data to understand urban flow.
\newblock In {\em Proceedings of the second international conference on IoT in
  urban space}, pages 1--7. ACM, 2016.

\bibitem{moran1950notes}
Patrick~AP Moran.
\newblock Notes on continuous stochastic phenomena.
\newblock {\em Biometrika}, 37(1/2):17--23, 1950.

\bibitem{o2014geographic}
David O'Sullivan and David Unwin.
\newblock {\em Geographic information analysis}.
\newblock John Wiley \& Sons, Hoboken, New Jersey, USA, 2014.

\bibitem{anselin1993moran}
Luc Anselin.
\newblock {\em The Moran scatterplot as an ESDA tool to assess local
  instability in spatial association}.
\newblock Regional Research Institute, West Virginia University, Morgantown,
  WV, 1993.

\bibitem{de2016death}
Marco De~Nadai, Jacopo Staiano, Roberto Larcher, Nicu Sebe, Daniele Quercia,
  and Bruno Lepri.
\newblock The death and life of great italian cities: a mobile phone data
  perspective.
\newblock In {\em Proceedings of the 25th international conference on world
  wide web}, pages 413--423. International World Wide Web Conferences Steering
  Committee, 2016.

\bibitem{graells2016sensing}
Eduardo Graells-Garrido, Oscar Peredo, and Jos{\'e} Garc{\'\i}a.
\newblock Sensing urban patterns with antenna mappings: the case of santiago,
  chile.
\newblock {\em Sensors}, 16(7):1098, 2016.

\bibitem{report}
{2017 Digital News Report}, reuters institute for the study of journalism and
  university of oxford.
\newblock
  \url{http://www.digitalnewsreport.org/survey/2017/chile-2017/#fn-6155-1}.
\newblock Accessed: 2019-02-26.

\bibitem{chan2015examining}
Michael Chan.
\newblock Examining the influences of news use patterns, motivations, and age
  cohort on mobile news use: The case of hong kong.
\newblock {\em Mobile Media \& Communication}, 3(2):179--195, 2015.

\end{thebibliography}
\end{document}